\documentclass{IEEEtran}
\usepackage{multirow}
\usepackage{graphicx}
\usepackage{amsmath}          
\usepackage{subfig}
\usepackage{authblk}
\usepackage{url}
\usepackage{array}
\usepackage{tabu}
\normalsize

\ifCLASSINFOpdf
\else
\fi

\usepackage{stfloats}
\usepackage{multicol}
\usepackage{pbox}
\usepackage{booktabs}
\usepackage{tabularx}
\usepackage[table,xcdraw]{xcolor}
\usepackage{xcolor}

\begin{document}
	\title{A Blockchain Policy and Charging Control Framework for Roaming in Cellular Networks}

		\author{
			\IEEEauthorblockN{Ahmed Refaey \IEEEauthorrefmark{1}\IEEEauthorrefmark{2}, Karim Hammad \IEEEauthorrefmark{1}, Sebastian Magierowski \IEEEauthorrefmark{1}, and Ekram Hossain\IEEEauthorrefmark{3} 
			}

			\IEEEauthorblockA{\IEEEauthorrefmark{1}York University, Toronto, Ontario, Canada.}\\
			\IEEEauthorblockA{\IEEEauthorrefmark{2} Manhattan College, Riverdale, New York, USA.}\\
			\IEEEauthorblockA{\IEEEauthorrefmark{3} 
				University of Manitoba, Manitoba, Canada.}}
\maketitle

\begin{abstract}
As a technology foundation of cryptocurrencies, blockchain enables decentralized peer-to-peer trading through consensus mechanisms without the involvement of a third party. Blockchain has been regarded as an auspicious technology for future cellular networks. It is able to provide solutions to problems related to mobile operators and user trust, embedded smart contracts, security concerns, pricing (e.g. for roaming), etc. When applying blockchain to cellular networks,
there are significant challenges in terms of deployment and application,  due to resource-constrained transactions. This article begins by introducing the basic concept of blockchain and then moves on to illustrate its benefits and limitations in the roaming system. Two models of roaming-based blockchain technologies are offered to show their suitability for cellular networks as opposed to traditional technology. Finally, potential issues and challenges of roaming-based blockchains are addressed and evaluated using the roaming use case in the EU.
\end{abstract}

%
\IEEEpeerreviewmaketitle


\section*{Introduction}
According to recent studies, a significant surge of mobile-connected devices per capita is expected to occur. Mobile network operators (MNOs) are anticipating that by the end of 2021 every person will own about 1.6 mobile-connected devices~\cite{1}. Triggered by their need to increase their network capacity, MNOs have embraced regional and national roaming, which allow for the sharing of resources. Moreover, the revenue of data roaming is expected to exceed US\$50 billion by 2021 \cite{2}.

Roaming offers both challenges and opportunities for operators, in areas such as technical feasibility, associated cost, and commercial implications of introducing regional, national, and international roaming for voice and data services. Although researchers and the industry have identified several roaming frameworks through authentication, authorization and accounting, still there remain critical issues associated with roaming. They are as follows \cite{3}-\cite{6}:\\

\begin{itemize}
	\item \textbf{Trust:} The illegal access to information related to the roaming user is not addressed in the existing architectures, which is considered a serious breach of the user protection Act. Precisely, the visitor network operator (VNO) may have differing policies than what is accepted by the users in their home network operator (HNO). This requires re-accepting the roaming privacy terms before providing any services.

	\item \textbf{Security:} The roaming user must be authenticated with the VNO through the HNO before gaining access. Through this procedure, the illegal users are thwarted from accessing the network. Indeed, there are many proposed user authentications and key exchange protocols for roaming services, however, these protocols only focus on mutual authentication between the roaming user and the VNO.

	\item \textbf{Scalability:} There is an increased interest by both MNOs and governing bodies to find ways to share resources efficiently. However, the peer-to-peer service agreement which is exercised by the MNOs is quite inefficient. Considering \textit{n} number of MNOs seeking roaming agreements, this requires $n(n-1)/2$ agreements in total, and $(n - 1) $ agreements for each MNO. This makes it impractical for a universal roaming service.

\end{itemize}

The cellular roaming scenario can be addressed by a blockchain approach. Precisely, the roaming procedures can be described as several smart contracts between the roaming user, HNO, and VNO. 
In this article, the blockchain approach is explored for roaming models in future cellular networks ~\cite{7}. In particular, agreements between subscribers and home network operators (HNO) are logged through the agency of smart contracts on a blockchain, such as Ethereum~\cite{7}.  These agreements associate a charging record with roaming (national and international) permissions and service instructions to be executed on an external database (i.e. the visitor network operator VNO). The motivating problem, which this article focuses on, is how to use blockchain to manage contracts between users and VNO, while maintaining the roaming agreements between MNOs. The proposed framework is based on developing trust between the parties to undergo trade, while maintaining full control over their respective user traffic in terms of quality-of-service (QoS), charging, and all the user's records (pre-approved privacy terms). Table \ref{Table 1} provides a comparison between the existing roaming model and the proposed blockchain based model.
Further, the contributions of this paper can be summarized as follows:
\begin{itemize}
	\item Proposing a blockchain architecture to the core mobile network.
	\item Introducing a mechanism that addresses the existing technological disadvantages of national and international roaming.
	\item Use Case: Evaluating the business implications of the European Union (EU) roaming charging model based on the blockchain from mobile operators' and users' point of views. 
\end{itemize}

The rest of this article is organized as follows: A background on the blockchain and smart contracts is provided and then the existing roaming architecture is presented with an explanation of its functionality. Afterwards, the proposed blockchain-based architecture for roaming is presented with its corresponding challenges. Finally, the new architecture is evaluated using a use case of the EU, before the article is concluded.

\newcommand{\cent}[1]{\begin{tabular}{l} #1 \end{tabular}}

\begin{table*}[]
	\caption{\label{Table 1}Roaming in the existing 3GPP standards \cite{8} versus blockchain}

	\centering

	\begin{tabular}
		{|c|p{0.32\linewidth} |p{0.32\linewidth}|}
		\hline
		\textbf{Aspect} & \multicolumn{1}{c|}{\textbf{Current 3GPP Standards}} & \multicolumn{1}{c|}{\textbf{Blockchain Approach}} \\ \hline
		 Ecosystems requirements  & \begin{tabular}{p{0.95\linewidth}}Needs commercial agreements including technical connections to support the IP signaling infrastructure \end{tabular}& \begin{tabular}{p{0.95\linewidth}}Needs predefined requirements to be implemented using an open-source software and commercial off-the-shelf (COTS) servers \end{tabular}\\ \hline
		Security & \begin{tabular}{p{0.95\linewidth}}IP border gateways and Diameter Edge Agents should be used for user paths and signaling can be utilized to securely connect roaming-related network interfaces to the roaming exchange network.\end{tabular} & \begin{tabular}{p{0.95\linewidth}}Offers a potentially more reliable network for roaming, free from security threats such as denial of service attacks that exist on public networks. \end{tabular}\\ \hline
		Performance and quality of experience (QoE)& \begin{tabular}{p{0.95\linewidth}}Partner operator networks cannot guarantee the same levels of operations and billing support, meaning the user experience can be negatively impacted.\end{tabular} & \begin{tabular}{p{0.95\linewidth}}Offers certainty about roaming data charges and avoid the potential for billshock when returning home. \end{tabular}\\ \hline
		Data and voice & \begin{tabular}{p{0.95\linewidth}}Requires managing multiple interfaces, routing calls (using circuit switch fallback or VoLTE), and allowing for redirections. \end{tabular} & \begin{tabular}{p{0.95\linewidth}}All data and voice calls are handled by the VNO and no routing  to the HNO is required.
		\end{tabular}\\ \hline
	\end{tabular}%

\end{table*}

\section*{Background}

The blockchain was originally designed to introduce a decentralized financial ledger ~\cite{9}. However, its paradigm was extended to provide several generalized frameworks for implementing decentralized compute resources~\cite{7}. In particular, the blockchain technology can be utilized to automate and track certain state transitions through the use of ``smart contracts"~\cite{10}. Thus, it allows alleviating the complexity of the modern charging control framework. Blockchain consists of three  main elements: 
\begin{itemize}
	\item {\em Transaction}: all the valuable information (e.g. ownership) can act as a transaction to be recorded in the blockchain. Therefore, it is not restricted to trading information. 
	\item {\em Block}: storage units to record transactions, which are created and broadcast by those users authorized by consensus mechanism. Also, each block is identified uniquely by its hash value, which is referenced by the block to come after it.
	\item {\em Chain}: the link between the blocks that creates a chain of blocks which is called a blockchain. The cost of attack and malicious modification increases exponentially as the blocks accumulate sequentially. 
\end{itemize}

Furthermore, the {\em consensus mechanism} in the blockchain, plays an indispensable role by resolving the trust concern. This is achieved by identifying the authorized parties to insert the next block into the blockchain. Several consensus mechanisms have been proposed for different types of blockchains such as Proof of Work (PoW) and Proof of Stake (PoS):

\begin{itemize}
	\item {\em PoW}: The core idea of PoW is the competition of computing power. In particular, the node performing the consensus mechanism  (called {\em miner}) uses its computing resources for the hashing operation to compete for the right to generate the new block with bonuses. The winner is the first one who obtains a hash value lower than the announced target.
	\item {\em PoS}: The coin age is used in PoS blockchain to avoid the high computational complexity of the hash operation. The coin age of an unspent transaction output (i.e. destination address and the amount of coin) is equal to its value multiplied by the time period after it was created. In PoS, a higher coin age will lead to a higher probability for the node to win the right of creating a new block, and in turn, the coin age would be consumed (reset as zero) when the owner wins. 
\end{itemize}

The blockchain concept has been widely adopted in various applications outside the financial ledger in the form of smart contracts. Anything expressible in digital form (e.g. computer program) can be seen as a smart contract. This computer program can use data from the blockchain records as inputs and then generate outputs. These outputs can be written to either the same blockchain or a separate one. In short, this computer program can digitally facilitate, verify, and enforce the contracts made between multiple parties on blockchain. Several attempts in the recent literature have used similar concepts for smart contracts, however, Ethereum~\cite{7} is considered the first full implementation of this concept. Table \ref{Table 2} provides a comparison between Ethereum and two other leading smart contract options. 
 
 \begin{table*}[!htbp]

		\caption{\label{Table 2}Differences among the three major smart contracts options}
\centering		
	\begin{tabular}{|c|p{0.36\linewidth}|p{0.36\linewidth}|}
		\hline
		\textbf{Name} & \multicolumn{1}{c|}{\textbf{Description}} & \multicolumn{1}{c|}{\textbf{Advantages}} \\ \hline
		Corda R3 \cite{11} & \begin{tabular}{p{0.95\linewidth}}\\A private permissioned distributed ledger technology (DLT)  where smart contracts are written in legal prose to mimic legally binding contracts. Participation is controlled by a trusted source and ledgers are private to those included on transactions.\end{tabular} & \begin{tabular}{p{0.95\linewidth}}\\1) Data is private to only those authorized to view it.  \\ 2) Consensus is two-fold: TX validity and uniqueness/ double spend prevention. No PoW or PoS. \\ 3) No need for incentives.  \\ 4) Provides higher TX/second. \\ 5) Smart contracts are associated with legal prose to account for high regulation environments.\end{tabular} \\ \hline
		Ethereum \cite{12} & \begin{tabular}{p{0.95\linewidth}}\\A public ledger blockchain that can be either permissioned or permissionless.  Consensus requires PoW, thus, miners and incentives are required (ETHER/ Tokens). Allows anyone to build smart contracts on it \end{tabular} & \begin{tabular}{p{0.95\linewidth}}\\1) Can be either private or public. \\ 2) Records are anonymized.  \\ 3) Rewards users with Ether or even customized reward schemes.  \\ 4) Versatile in a wide variety of use cases.\end{tabular} \\ \hline
		Hyperledger Fabric \cite{13} & \begin{tabular}{p{0.95\linewidth}}\\Private permissioned blockchain with unique roles. Nodes can be either clients, orders or peers. Clients invoke transactions, peers maintain the ledger, and the orderer orders the new transactions. A special type of peer known as endorsers check the provision of signatures on a transaction \end{tabular} & \begin{tabular}{p{0.95\linewidth}}\\1) Data is private to only those 'subscribed' to a channel. \\ 2) Finer grain control over consensus. (\emph{i.e.} it can use Byzantine fault tolerance (BFT) consensus algorithm for faster performance and higher scalability than PoW or PoS blockchains). \\ 3) Can be used with or without incentive tokens.\end{tabular} \\ \hline
	\end{tabular}%
\end{table*}

\section*{Existing Roaming Models}

The main objective of a mobile roaming service is to provide a service for any user with a valid international mobile subscriber identity (IMSI). In principle,  subscribers of a specific service might go outside the pre-scheduled coverage zone by their home network operator (HNO). Consequently, an indirect connection for accessing such services is entailed via visitor network operators (VNOs). This is referred to as regional or national roaming which requires a service agreement to be set up between the home and visited networks. There are three parties involved in such a scenario: a roaming subscriber, the subscriber's HNO, and the VNO.  In addition to the roaming service agreement, an essential authentication process concurring with the roaming registration procedures must be accomplished to ensure legitimate access to the visitor network resources and proper billing to roaming service-subscribed users~\cite{14}. One known key challenge introduced by the authentication process pertains to its latency. As defined by the IEEE standard ~\cite{15}, to maintain persistent connectivity for roaming users the hand-over latency has to be bounded by a delay constraint of 50 msec. Thus, highly efficient authentication mechanisms are needed especially with today’s highly dense cellular networks. Further, these aforementioned procedures should be done with all legacy trust mechanisms which require a complex and costly infrastructure to maintain.

In general, there are two models of roaming as follows:
\begin{enumerate}
	\item {\it International roaming}: It is a service through which a user of a given MNO (i.e. HNO) can obtain service from a MNO (i.e. VNO) of another country which is subject to inter-MNO agreements.
	\item {\it National roaming}: It is a service through which a user of a given MNO (i.e. HNO) can obtain service from an alternate MNO (i.e. VNO) of the same country, on a regional or everywhere basis. This roaming type depends on the agreement between the HNO of the requesting user and the VNO while being independent from user subscription arrangements.
\end{enumerate}

\begin{figure*}[htbp]
	\centering
	{\includegraphics[width=1\textwidth,height=7cm]{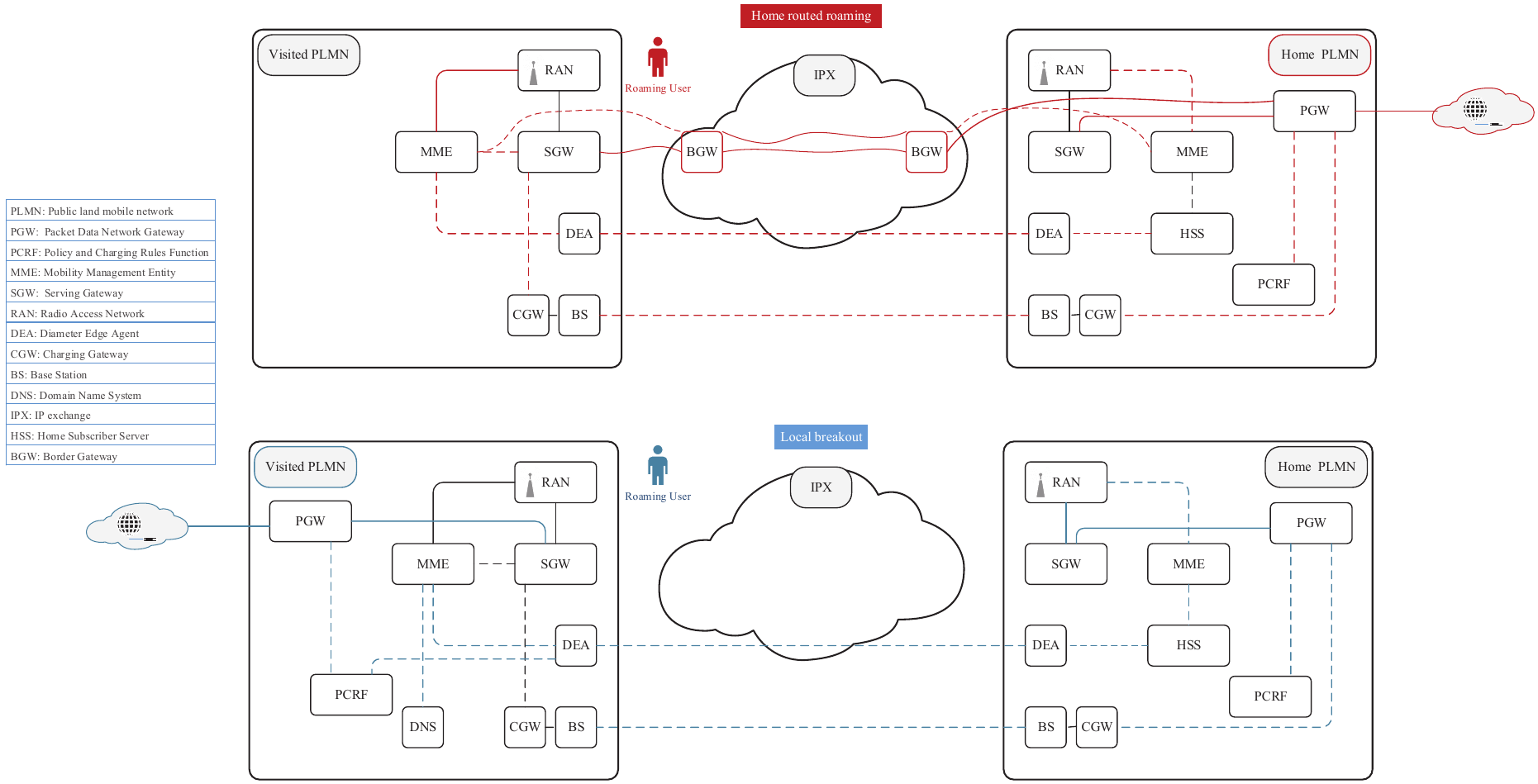}}
	\caption{Block diagram of the home routed and local breakout roaming models.}
	\label{Drawing12}
\end{figure*}

Regardless of the different interfaces between home and visited MNOs, based on the 3GPP model, these two roaming services utilize the same network architecture. Currently, the 3GPP standards support two roaming models as described in the following subsections~\cite{8} and shown in Fig. \ref{Drawing12}:

\begin{enumerate}
	\item {\em Home-routed roaming architecture}: users' data traffic is serviced by their home network and gives the network operator more control over the users' traffic.
	\item {\em Local breakout architecture}: users' data is serviced by the network they are visiting. Therefore, it delivers more efficient routing in terms of bandwidth and latency~\cite{8}.
\end{enumerate} 

\section*{Proposed Blockchain Roaming Model}

In terms of roaming, there are many potential opportunities for operators to use the blockchain concept. In particular, by ensuring that proper agreements and users' records or users' transaction records are efficiently 
shared/maintained among fellow operators, it becomes possible to generate new revenues through innovative roaming service offerings. In addition, the deployment of such common architecture across operators will define a new universal roaming platform on which existing and new operators can better serve their clients everywhere with minimal to no changes in their existing infrastructure.

Regardless of the roaming model chosen, a blockchain can enable a unified and flexible service offering for the roaming users. As shown in Fig. \ref{Drawing14}, this can be achieved among different core network operators using smart contracts among themselves and predefining their relative share of roaming users. Consequently, the user can register with any operator given that the operator is among the network sharing partners. The selection of an operator among those network sharing partners should be done manually beforehand to allow the accepting or rejecting of the roaming terms for the user (e.g. for roaming privacy and charging tariff). It is noteworthy that, the charging tariff in this system will be more transparent to the user than in the existing one. For example, if the cryptocurrency is used and the user pays half cryptocurrency in the HNO, it will be easier for the user to decide when the VNO offers the same service by quarter cryptocurrency.

The specific agreement is represented by a state-object, which can be considered the main building block in this system. Below, are three main tools to achieve a blockchain roaming-based consensus:
\begin{itemize}
	\item {\em Smart contract}: Logic which specifies constraints that ensure that the state transitions are valid according to the roaming agreement among the operators.
	\item {\em Transaction}: Showing the transition state-objects through a lifecycle of the roaming user.
	\item {\em Flow framework}: A component to simplify coordinating actions without a central controller among the sharing network partners.
\end{itemize}

\subsection{Blockchain System Model}

The smart contracts can be referred to as a generalized state-machine of the blockchain based on cryptographically-secured transactions. Precisely, the blockchain paradigm is implementing decentralized compute resources whereas each MNO as a computing resource node can be considered as a state-machine entity. A state transition is determined to be valid by the MNO nodes encoding logic. This occurs for every newly generated state-machine and then uploaded onto the blockchain. In this blockchain, the data permissions (such as ownership and view) is presented as block content and shared by members of the network sharing partners. As a consequence, the blocks account a series of valid transactions. These transactions, utilize the previous block’s state to incrementally transform the state-machine into its ultimate state. All nodes engaged in the system receive this information through the PoW consensus algorithm. Furthermore, the PoW secures the state-machines’ state and transitioning logic against any changes.  
It is noteworthy that MNO nodes are capable to query the state-machines at any time and the obtained results would be accepted by the entire network. 

\begin{figure*}[htbp]
	\centering
	{\includegraphics[width=1\textwidth,height=8cm]{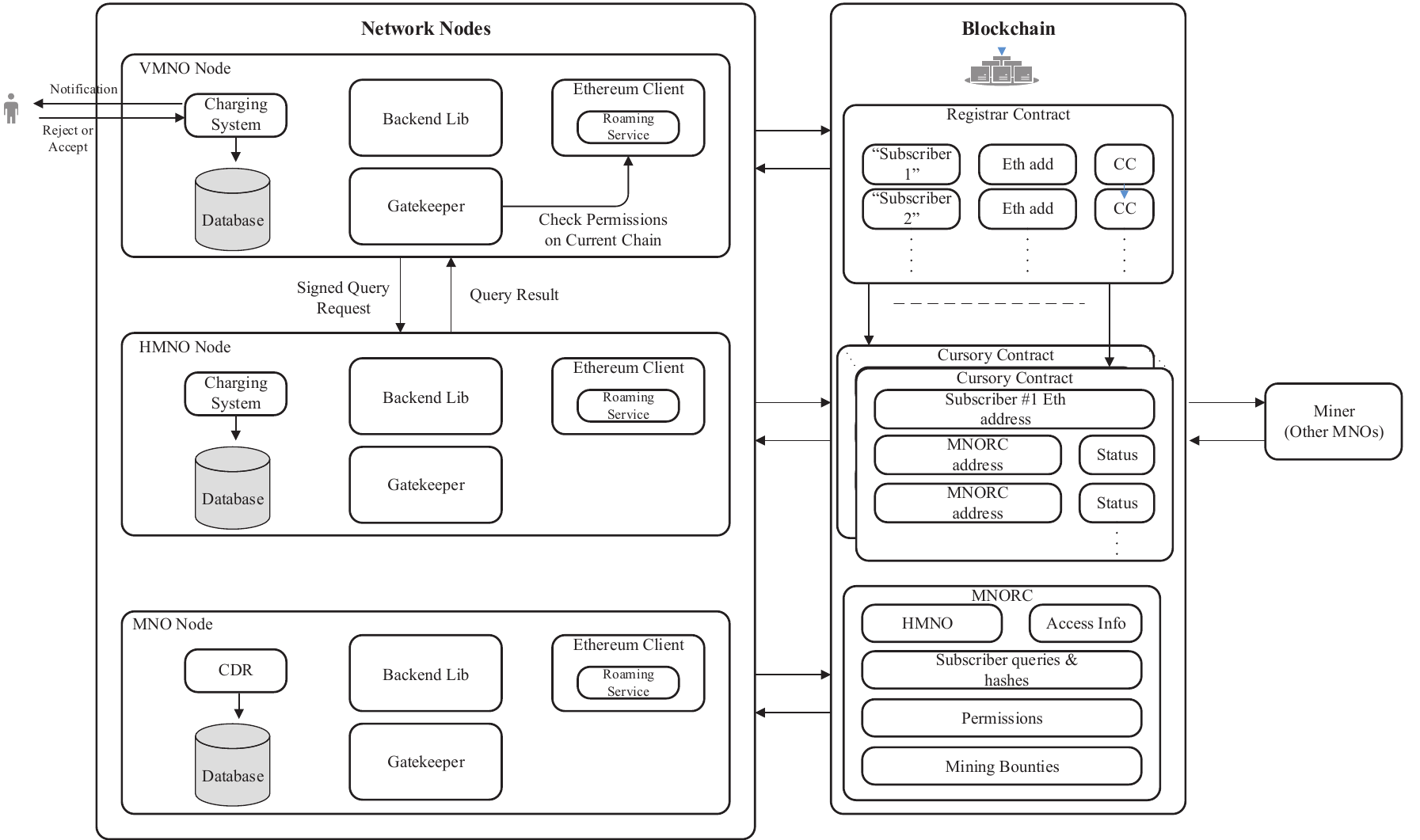}}
	\caption{Architecture of the proposed blockchain roaming model.}
	\label{Drawing14}
\end{figure*}

In the proposed system, the Ethereum's smart contracts are used to establish smart representations for the users roaming records which are shared among operators on the shared network. These contracts contain parameters such as ownership, permissions and data integrity which are recorded in the form of metadata. Also, all legitimate transactions are cryptographically signed in order to manage these records. Depending on these transactions, the contract's state-transition functions will be able to enforce data alteration and carry out any policies. These policies are represented computationally and can implement several rules which administer a specific roaming record. In particular, a policy can enforce the granting of roaming permissions after a separate transaction sent from the VNO verifies the user's acceptance.

As shown in Fig.~\ref{Drawing14}, there are three types of contracts which are implemented in order to navigate the expected volume of record representations. The details for these contracts are provided in the following subsections.

\subsubsection{Mobile Network Operator Contract (MNOC)} This contract is issued between two operators when a user attempts to register in a VNO. The MNOC defines a collection of data pointers that identify the records held by the HNO as well as the associated access permissions. Each of these data pointers consists of query string which retrieves the user data when executed on the HNO's database. The hash of the data subset is append to the query string to prevent any data alteration at the source. Also, more information such as the QoS attributes could be accessible to the VNOs through the HNO's database. It is noteworthy that, a design based on generic strings can simplify the interface with the different string queried database implementation for each operator on the network.

\subsubsection{Cursory Contract (CC)} This contract depicts all the users' past and current interactions with any operator in the system by holding a reference list to all Mobile Network Operator Contracts. 

\begin{itemize}
	
	\item {\em Users}: their CC would be populated with references to all operators they have interacted with.
	
	\item {\em HNOs}: would have references to their users and all VNOs with which their users have established authorized data sharing.
	
\end{itemize}

The CC goes on the distributed network to add functionality such as backup, restoring, and enable user notifications. Also, to further establish a user acceptance, a status variable is stored with each roaming relationship to indicate whether this relationship is recently formed or awaiting updates. The HNO has the capacity to set a roaming status in their users' CC while updating records or while creating a new roaming activity. Hence, the user can poll their CC and be notified in the case a new roaming activity is suggested which is crucial in enhancing the QoE through providing additional national roaming options. Ultimately, users can accept or reject the suggested roaming activity.

\subsubsection{Registrar Contract} Users and operators identification (ID) strings are mapped to their Ethereum address identity with this global contract. These strings are equivalent to a public key and allow the users to use a unique serial number ( \emph{i.e.,} the Integrated Circuit Card Identifier (ICCID)). In general, if policies are coded into this contract they regulate both new identities being registered and changes in the mapping of existing ones. Consequently, identity registration can be limited to participant operators. Furthermore, identity strings are mapped to an address located on the blockchain by this contract.


\section*{System Description}

Agreements between subscribers and HNO are logged through the agency of smart contracts on the Ethereum blockchain. These agreements associate a charging record with roaming permissions and data retrieval instructions for execution on an external database (i.e. VNO). Once the agreement is settled, the remaining step for the MNO to allow roaming is having the HNO's permission to access the subscriber’s charging record.  Any of the MNOs on the network can be designated as a VNO by obtaining access to the charging record of the subscriber. Consequently, a temporary agreement will be issued between the VNO, HNO, and U. This new temporary agreement will allow the VNO to make additions to the subscriber record. This operation will include accepting the new additional terms from the subscriber and notifying the HNO.
Charging between the MNOs is done inside the network using their own cryptocurrency. Furthermore, implementing a universal charging system means applying a conversion rate between subscribers' currency and the MNO cryptocurrency, which the subscriber will be notified of by the VNO. Consequently, the subscriber will have the option to either accept or reject the additional temporary terms and charges imposed by the VNO. This scenario will be helpful in the case of international roaming, however, in the case of regional or national roaming charges will be excluded in the temporary terms but handled between the MNOs.

As shown in Fig. \ref{Drawing14}, the proposed system orchestration is based on four software components which create a coherent and
distributed system as follows: 
  
\begin{itemize}
	\item {\em Backend Library}: Facilitates the system's operations by abstracting the communications with the blockchain.
	\item {\em Ethereum Client}: Implements the required functionality to join the blockchain network such as encoding and sending transactions, maintaining a local copy of the blockchain, and connecting to the P2P network.
	\item {\em Gatekeeper}: This is a database which implements an off-chain access interface to the MNO local database as well as listening to query requests from other MNOs by running a server.
	\item {\em Charging System}: It renders data from a local database for reviewing, and presents a notification to the roaming user to either accept or reject the roaming terms. 
\end{itemize}

By implementing the modular interoperability protocol, any new operator can participate in the network, as the previous section has outlined. Also, by following the CC, missing data can be recovered from the network on demand.

\section*{Model and Performance Analysis } 
Since 2017, the European parliament enforced free roaming in order to achieve a unified digital market. In fact, this goal has had an impact on both users and MNOs through the pricing strategies and transit payments between MNOs in different EU countries. As a consequence, the model in ~\cite{16} is adopted to analyze the depending parameters from the users, VMOs and HNOs perspective.  This can be considered a practical example for the proposed framework with the ultimate goal of having a global digital market based on the Blockchain approach.

From the MNO's perspective, revenue is the performance metric to optimize, which can be expressed in a high-level, for country \textit{i}, as:
\begin{equation}\label{equ:Rev}
\begin{array}{l}
{R_i} = \,{\rm{domestic\, usage\, revenue\, for\, }}MN{O_i}\,{\rm{  + }}\\
{\rm{\,\,\,\,\,\,\,\,\,\,\,\,\,\,\,\,       roaming\, usage\, revenue\, for\, }}MN{O_i}\,\,{\rm{  + }}\\
{\rm{\,\,\,\,\,\,\,\,\,\,\,\,\,\,\,\,       transit\, usage\, revenue\, for\, }}MN{O_i}\,\,\,\,\,\,{\rm{  -  }}\\
{\rm{\,\,\,\,\,\,\,\,\,\,\,\,\,\,\,\,       transit\, usage\, cost\, incurred\,  by\, }}MN{O_i}{\rm{ }}
\end{array}
\end{equation}
where the four summands reflect three revenue components and one cost component. To simplify the understanding of \eqref{equ:Rev}, each component is defined as follows: 
\begin{itemize}
\item The \textbf{domestic usage revenue} is the flat-rate subscription payment paid by domestic customers to their HNO for domestic use. This revenue component is proportional to ${m_i}\,{p_i}\,\frac{1}{{{\lambda _i}\,{{\overline \theta  }_i}}}$, where $m_i$ is the total number of potential users in country \textit{i}, $p_i$ is the flat-rate subscription price for the domestic service of the MNO in country \textit{i}, $\lambda_i$ is the wealth parameter of country $i$ and defined as the rate of the (assumed) exponential distribution for the users' willingness-to-pay, $\theta_i$ (a measure in monetary units per month) in country $i$.  Finally, $\overline \theta_i$ is the minimum willingness-to-pay value to consider a user as a potential subscriber with the MNO in $i$.  Given the assumed exponential distribution of $\theta_i$, the country's wealth parameter $\lambda_i$ is the reciprocal of the country's average wealth.
  
\item The \textbf{roaming usage revenue} is the payments from domestic users to their HNO for roaming use. This revenue component is proportional to $m_i c_i\,\frac{1}{{{\lambda _i}\,{\overline \theta_i}}}$, where $c_i$ is the fixed per-volume price for the roaming service of the MNO in country \textit{i}.
\item The \textbf{transit usage revenue} is the payments (transit fees) to MNO$_i$ for the roaming costs incurred by foreign roaming users for having their traffic traversing MNO$_i$'s infrastructure. This revenue component is proportional to ${m_j}\,{t_i}\,\frac{1}{{{\lambda _j}\,{{\overline \theta  }_j}}},\,\forall j \ne i$, where $t_i$ is the transit price decided by $\rm{MNO_i}$ to charge counterpart MNOs in other countries for using its network infrastructure.

\item An Alternative to transit usage revenue is the \textbf{transit usage cost} which is the payment from $\rm{MNO_i}$ for the roaming costs incurred by their domestic customers for roaming on foreign networks. This cost component is proportional to ${m_i}\,{t_j}\,\frac{1}{{{\lambda _i}\,{{\overline \theta  }_i}}},\,\forall j \ne i$. 
\end{itemize}

Conversely, from the users' perspective, the consumer surplus (CS) is used to measure the aggregated satisfaction level of all subscribed users, across all countries, and can be expressed as:
\begin{align}\label{equ:CS}
\begin{array}{l}
{\rm{CS}} = \,{\rm{net\, utility\, for\, domestic\, usage\,  + }}\\
{\rm{\,\,\,\,\,\,\,\,\,\,\,\,\,\,\,\,\,net\, utility\, for\, roaming\, usage}}
\end{array}
\end{align}
where both summands represent, in order from left, the users' satisfaction level (and essentially the users' subscription decision) about the MNO's flat subscription price for the domestic service and the fixed per-volume price for the roaming service, respectively. In particular, the utility for domestic usage is equal to $\theta _i{-}p_i$ whereas the utility for roaming usage is proportional to ${({\theta _i} - {c_i})^2}$.

The values of $p_i$ and $c_i$ are selected to maximize the operator's revenue expressed in (\ref{equ:Rev}), whereas $t_i$ is decided by two possible schemes. The first scheme involves leaving the transit price decision to the regulator to maximize the CS. The authors in~\cite{16} have concluded that zero transit prices (i.e. $t_i$= 0, $\forall i$) is likely to be the optimal scenario where the CS is maximized. The second scheme, alternatively, finds the transit price at which a Nash equilibrium for a non-cooperative game is reached between MNOs.

The performance comparison between the traditional roaming system and the proposed blockchain-based roaming system is conducted by evaluating equations (\ref{equ:Rev}) and (\ref{equ:CS}) for both systems. On the one hand, for the traditional roaming system, both equations are analyzed for the paid roaming and free transit case as in \cite{16} (i.e. the transit revenue and cost components in (\ref{equ:Rev}) are equal to 0). On the other hand, the evaluation for both equations in the case of our proposed system is discussed in the following two paragraphs while explaining its relative performance. To ensure fair comparison between both systems, we set the pricing model (i.e. explained in (\ref{equ:Rev}) and (\ref{equ:CS})) parameters to similar values as taken in \cite{16} (i.e. $m_1= 1$, $m_2= 2$ and $\lambda_2= 1$). As such, the impact of the wealth parameter for operator 1 (i.e. $\lambda_1$), as well as the rest of the roaming model parameters, on the operators' revenue and consumer surplus is depicted in Tables \ref{fig:op_rev} and \ref{fig:cs}, respectively.

In Table \ref{fig:op_rev}, we notice that increasing $\lambda_1$ the operator's revenue for MNO$_1$, $R_1$, becomes negatively impacted in case of both traditional and proposed roaming models. This pertains to the fact that the smaller the average wealth the smaller the number of potential subscribers for the MNO service. In addition, MNO$_1$ is able to gain higher revenues when implementing the proposed blockchain-based model than in the case of traditional roaming. This can be justified by the higher roaming revenue gained from MNO$_2$ subscribers roaming to MNO$_1$. In other words, in the case of traditional roaming, the roaming revenue for MNO$_1$ comes from their own subscribers which decreases by increasing $\lambda_1$. However, in the case of blockchain roaming, the roaming revenue for MNO$_1$ comes from MNO$_2$ customers roaming to MNO$_1$. keeping in mind that $\lambda_2$ is constant and equal to 1 and $m_2>m_1$, hence, the roaming revenue from MNO$_2$ roaming subscribers (in case of blockchain roaming) is relatively greater than that from MNO$_1$ roaming subscribers (in case of traditional roaming) with increasing $\lambda_1$. In essence, in the blockchain-based roaming system, the roaming process is highly flexible such that the user seamlessly receives the required service from the VNO by directly paying for it under the smart-contract authorization system without intervention from the HNO. On the other hand, the revenue for the traditional roaming MNO$_2$ is not affected by the average wealth in country 1. However, the proposed blockchain roaming improves MNO$_2$ revenue only when $\lambda_1<\lambda_2$ (i.e. at $\lambda_1= 0.5$) which implies that MNO$_1$ subscribers roaming to MNO$_2$ are wealthier than MNO$_2$ domestic subscribers and would contribute more to the roaming usage revenue for MNO$_2$ (as illustrated in (\ref{equ:Rev})) compared to the traditional roaming case. In general, improving $R_1$ on the expense of $R_2$ when implementing the proposed blockchain roaming system is consistent with the relative difference between $\lambda_1$ and $\lambda_2$.

%

\begin{table}[ht]
\caption{\label{fig:op_rev} Operator's revenue analysis}
\centering
\begin{tabular}{|p{1.4cm}|p{1cm}|p{1cm}|p{1cm}|p{1cm}|p{1cm}|}
\hline   
\multicolumn{6}{|c|}{\includegraphics[scale=0.33]{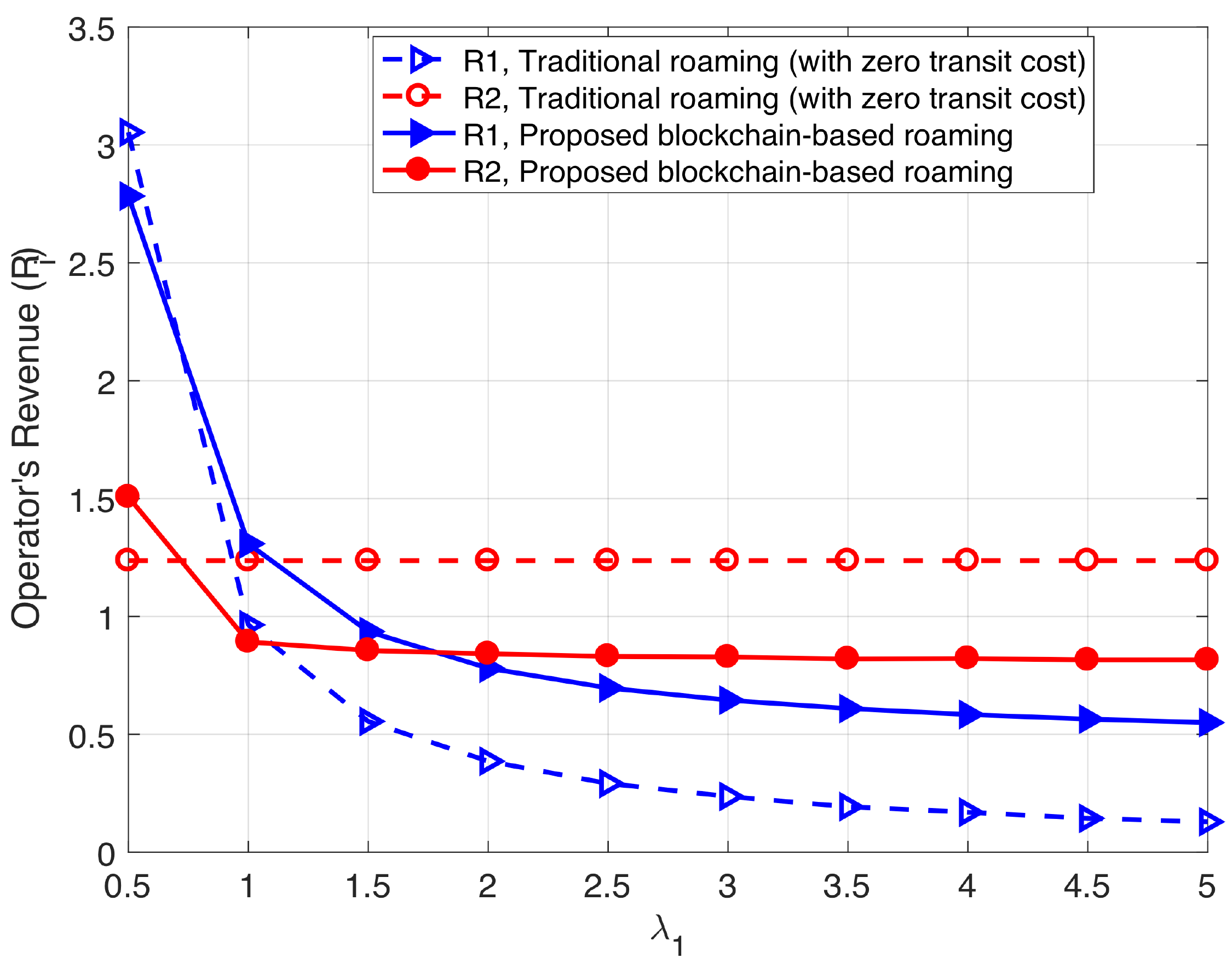}} \\ \hline
\textbf{Parameter} & $p_i$ & $c_i$ & $t_i$ & $\lambda_i$ & $\theta_i$\\ [1ex] \hline
\textbf{Operator's revenue} & Increases & Increases & Incr./Decr. & Decreases & Increases \\ \hline
\end{tabular}
\end{table}

From the users' perspective, the effect of $\lambda_1$ on the consumer surplus is illustrated in Table \ref{fig:cs}. The results illustrate that the CS drops with increasing $\lambda_1$ due to the fact that the larger the $\lambda_1$ the larger the density of small values for $\theta_1$, and hence, inversely affects the CS. As highlighted in the previous paragraph, due to the flexible nature of the proposed blockchain-based roaming which allows roaming subscribers to pay for the service on-the-go with the selected VNO, the traditional roaming charge for the HNO drops to 0. With this in mind, the roaming usage utility in (\ref{equ:CS}) is maximized, and hence, the aggregated CS as depicted in Table \ref{fig:cs}. 


\begin{table}[ht]
\caption{\label{fig:cs} Consumer surplus analysis}
\centering
\begin{tabular}{|p{1.4cm}|p{1cm}|p{1cm}|p{1cm}|p{1cm}|p{1cm}|}
\hline   
\multicolumn{6}{|c|}{\includegraphics[scale=0.33]{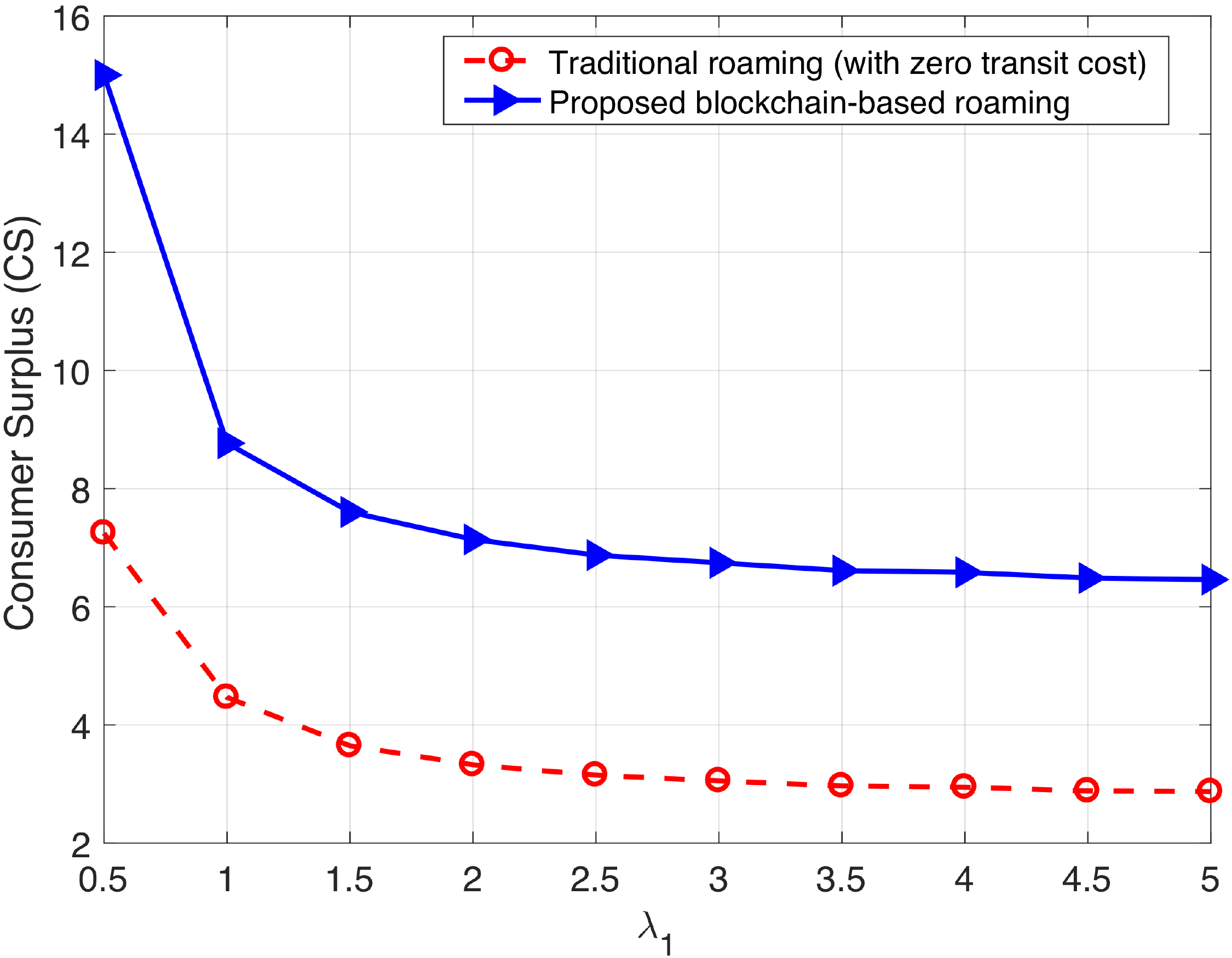}} \\ \hline
\textbf{Parameter} & $p_i$ & $c_i$ & $t_i$ & $\lambda_i$ & $\theta_i$\\ [1ex] \hline
\textbf{Consumer surplus} & Decreases & Decreases & Decreases & Decreases & Increases \\ \hline
\end{tabular}
\end{table}


\section*{Conclusion and Future Directions}

The blockchain vision for telecommunication networks enables multiple applications and services to run across a common layer of identity, business logic, and governance. Operators can deploy one or multiple blockchain infrastructures which can transact for various purposes while maintaining fortified privacy. In this article, the concept of blockchain
and the benefits of integrating it into the Roaming systems have been presented. The main ideas and basic roaming procedures based on smart contacts have been illustrated. Further, a comparison  among the state-of-the-art smart contract-based distributed ledgers have been provided in terms of their main characteristics. A case study for the proposed model has also been presented.  From the operators' perspective, the MNOs will still be able to achieve a reasonable revenue while providing ubiquitous connectivity for numerous mobile devices with minimal changes in their infrastructure which will drastically reduce the OPEX. From the users' perspective, the user will be able to obtain the maximum QoE through the capability of choosing the service and accepting the cost.

\bibliographystyle{IEEEtran}
\bibliography{IEEEabrv}

\begin{thebibliography}{10}
\providecommand{\url}[1]{#1}
\csname url@samestyle\endcsname
\providecommand{\newblock}{\relax}
\providecommand{\bibinfo}[2]{#2}
\providecommand{\BIBentrySTDinterwordspacing}{\spaceskip=0pt\relax}
\providecommand{\BIBentryALTinterwordstretchfactor}{4}
\providecommand{\BIBentryALTinterwordspacing}{\spaceskip=\fontdimen2\font plus
\BIBentryALTinterwordstretchfactor\fontdimen3\font minus
  \fontdimen4\font\relax}
\providecommand{\BIBforeignlanguage}[2]{{%
\expandafter\ifx\csname l@#1\endcsname\relax
\typeout{** WARNING: IEEEtran.bst: No hyphenation pattern has been}%
\typeout{** loaded for the language `#1'. Using the pattern for}%
\typeout{** the default language instead.}%
\else
\language=\csname l@#1\endcsname
\fi
#2}}
\providecommand{\BIBdecl}{\relax}
\BIBdecl

\bibitem{1}
\BIBentryALTinterwordspacing
Anon., ``Cisco visual networking index: Global mobile data traffic forecast
  update, 2016--2021,'' Feb. 2017. [Online]. Available:
  \url{https://www.cisco.com/c/en/us/solutions/collateral/service-provider/visual-networking-index-vni/mobile-white-paper-c11-520862.pdf}
\BIBentrySTDinterwordspacing

\bibitem{2}
I.~T.~. Media, ``{LTE Roaming: Global Market Status Drivers for Growth
  Deployment Plans},'' TATA Communications, Tech. Rep., 07 2013.

\bibitem{3}
W.~. {Liao} and Y.~. {Chen}, ``Supporting vertical handover between universal
  mobile telecommunications system and wireless lan for real-time services,''
  \emph{IET Communications}, vol.~2, no.~1, pp. 75--81, January 2008.

\bibitem{6}
M.~{Steeg}, N.~J. {Gomes}, A.~A. {Juarez}, M.~{Kosciesza}, M.~{Lange},
  Y.~{Leiba}, H.~{Mano}, H.~{Murata}, M.~{Szczesny}, and A.~{Stohr}, ``Public
  field trial of a multi-rat (60 ghz 5g/ lte/wifi) mobile network,'' \emph{IEEE
  Wireless Communications}, vol.~25, no.~5, pp. 38--46, October 2018.

\bibitem{7}
\BIBentryALTinterwordspacing
G.~Wood, ``Ethereum: A secure decentralised generalised transaction ledger
  eip-150 revision (759dccd - 2017-08-07),'' 2017, accessed: 2018-01-03.
  [Online]. Available: \url{https://ethereum.github.io/yellowpaper/paper.pdf}
\BIBentrySTDinterwordspacing

\bibitem{8}
\emph{Study on roaming architecture for voice over IP Multimedia Subsystem
  (IMS) with local breakout}, 3GPP Specification, 4 2013, release 11.

\bibitem{9}
S.~Nakamoto, ``Bitcoin: A peer-to-peer electronic cash system,''
  \emph{Cryptography Mailing list at https://metzdowd.com}, 03 2009.

\bibitem{10}
\BIBentryALTinterwordspacing
N.~Szabo. (1994) Smart contracts. [Online]. Available:
  \url{https://web.archive.org/web/20011102030833/http://szabo.best.vwh.net:80/smart.contracts.html}
\BIBentrySTDinterwordspacing

\bibitem{11}
I.~G. M.~H. Richard Gendal~Brown, James~Carlyle, ``{Corda: An Introduction},''
  Corda.net, Tech. Rep., 01 2016.

\bibitem{12}
V.~Buterin, ``{A Next-Generation Smart Contract and Decentralized Application
  Platform},'' Ethereum.org, Tech. Rep., 02 2015.

\bibitem{13}
T.~B. et~al., ``{An Introduction to Hyperledger},'' Hyperledger.org, Tech.
  Rep., 07 2018.

\bibitem{14}
K.~{Park}, Y.~{Park}, Y.~{Park}, A.~{Goutham Reddy}, and A.~K. {Das},
  ``Provably secure and efficient authentication protocol for roaming service
  in global mobility networks,'' \emph{IEEE Access}, vol.~5, pp.
  25\,110--25\,125, 2017.

\bibitem{15}
C.~Cox, \emph{An Introduction to LTE: LTE, LTE-Advanced, SAE, VoLTE and 4G
  Mobile Communications}, 2nd~ed.\hskip 1em plus 0.5em minus 0.4em\relax Wiley
  Publishing, 2014.

\bibitem{16}
\BIBentryALTinterwordspacing
P.~Maill{\'{e}} and B.~Tuffin, ``How does imposing free roaming in {EU} impact
  users and isps' relations?'' in \emph{8th International Conference on the
  Network of the Future, {NOF} 2017, London, United Kingdom, November 22-24,
  2017}, 2017, pp. 27--32. [Online]. Available:
  \url{https://doi.org/10.1109/NOF.2017.8251216}
\BIBentrySTDinterwordspacing

\end{thebibliography}

\end{document}